\documentclass[12pt]{article}
\usepackage{amssymb,amsmath}
\usepackage{latexsym}
\usepackage{graphicx}
\usepackage{cite}
\usepackage[usenames]{color}
\usepackage[linktocpage=true]{hyperref}

\makeatletter
\@addtoreset{equation}{section}
\makeatother

\def\Neql#1{{\cal N}\!=\!{#1}}
\def\nBPS#1{$\frac{1}{#1}$-BPS}
\def\oneone{\rlap 1\mkern4mu{\rm l}}
\def\coeff#1#2{\relax{\textstyle {#1 \over #2}}\displaystyle}

\def\IC{\mathbb{C}}
\def\IP{\mathbb{P}}
\def\IR{\mathbb{R}}
\def\ZZ{\mathbb{Z}}

\def\cB{{\cal B}}
\def\cD{{\cal D}}

\def\cN{{\cal N}}

\def\cR{{\cal R}}

\definecolor{cardinal}{rgb}{0.6,0,0}
\definecolor{darkgreen}{rgb}{0,0.4,0}
\definecolor{purple}{rgb}{0.5, 0, 0.5}
\definecolor{golden}{rgb}{0.92, 0.7, 0}
\definecolor{midnight}{rgb}{0, 0, 0.5}
\definecolor{darkblue}{rgb}{0, 0, 0.8}

\begin{document}
\begin{titlepage}

\begin{flushright}
\end{flushright}

\begin{center}

\bigskip
\bigskip
\centerline{\Large \bf  Supersymmetry and Wrapped Branes in Microstate Geometries}
\date{\today}
\bigskip
\bigskip
\centerline{{\bf Alexander Tyukov$^1$ and Nicholas P. Warner$^{1,2}$ }}
\bigskip
\centerline{$^1$ Department of Physics and Astronomy}
\centerline{and $^2$Department of Mathematics}
\centerline{University of Southern California} \centerline{Los
Angeles, CA 90089, USA}
\bigskip
\bigskip

\begin{abstract}
\noindent

\noindent We consider the supergravity back-reaction of M2 branes wrapping around the space-time cycles in \nBPS{8} microstate geometries.  We show that such brane wrappings will generically break all the supersymmetries.  In particular, all the supersymmetries will be broken if there are such wrapped branes but the net charge of the wrapped branes is zero.  We show that if M2 branes wrap a single cycle, or if they wrap a several of co-linear cycles with the same orientation, then the solution will be \nBPS{16}, having two supersymmetries.  We comment on how these results relate to using W-branes to understand the microstate structure of  \nBPS{8} black holes.

\end{abstract}

\end{center}

\end{titlepage}

\tableofcontents

\section{Introduction}
\label{sec:intro}

There are very large families of solitonic solutions to supergravity that appear, from far away, to be like black holes and yet cap off in smooth geometry as one approaches the core of the solution \cite{Bena:2005va,Berglund:2005vb,Bena:2007kg}. The vast majority of the known  solitons are supersymmetric and, for those that have the charge and angular momenta corresponding to black holes with macroscopic horizons, the core of the solution asymptotes to an arbitrarily long, possibly rotating, $AdS_p \times S^q$ throat \cite{Bena:2006kb,Bena:2007qc}.  The geometry then caps off just above where the black-hole horizon would be.  Classically, the depth of the throat, and hence the red-shift between the cap and infinity, is a freely choosable parameter determined by the moduli of the soliton.  While the capping-off was long known to be the result of a geometric transition in which non-trivial homology cycles blow up, supported by magnetic flux, it was only relatively recently that it was shown that the non-trivial topology is an absolutely essential ingredient if one is to have smooth solitons in supergravity \cite{Gibbons:2013tqa}. 

One of the primary motivations for studying such supergravity solitons is the microstate geometry program in which the solitonic geometries are related to the microstate structure of  black holes with the same asymptotic charges.   For BPS black holes with $AdS$ throats, this can be made very precise using holographic field theory and, in particular,  IIB microstate geometries, and fluctuations around them, can be mapped directly onto  states in the D1-D5 CFT. (For some recent results, see  \cite{Giusto:2013bda,Bena:2014qxa,Giusto:2015dfa,Bena:2015bea,Bena:2016ypk}.)  Deep, scaling microstate geometries can access states in the lowest-energy sectors of the D1-D5 CFT that contain much of the microstate structure. Indeed, once the classical moduli space of these geometries is quantized, the depth of the throat and the redshift are limited in precisely such a manner as to holographically reproduce the energy gap of the maximally twisted sector of the CFT \cite{Bena:2006kb,Bena:2007qc,deBoer:2008zn,deBoer:2009un}.

A major focus of the microstate geometry program has been to study of the BPS fluctuation spectrum in supergravity and to determine the dual CFT states. The goal of this work has been to see the extent to which such fluctuations can sample the microstate structure and perhaps give a semi-classical description of the black-hole thermodynamics.  While there has been a lot of progress on this in recent years \cite{Giusto:2013bda,Bena:2014qxa,Giusto:2015dfa,Bena:2015bea,Bena:2016ypk}, it remains unclear as to whether such techniques will yet give a semi-classical description of the entropy. The challenge that remains is how supergravity can be used to access a rich variety of states within the highly twisted sectors of the CFT.  There are quite a few ideas as to how this might be achieved but, as present, the required computations are extremely challenging.  

Independent of the fluctuation story, there is the broader application of microstate geometries as backgrounds on which to study string theory and to see if one can use intrinsically stringy excitations of these geometries to access the microstate structure of black holes. One of the most interesting and promising  approaches to this is the study of  branes wrapping the non-trivial space-time homology cycles that support the microstate geometry.

Perhaps the first motivation for studying such wrapped branes is the black-hole `deconstruction' story   \cite{Denef:2007yt,Gimon:2007mha}.  This starts with a fully-back-reacted geometry consisting of a D6-$\overline{\rm D6}$ pair with fluxes that induce  with D4-D2-D0 charges.   The scaling geometry is then $AdS_3 \times S^2$.   A  gas of D0 branes is added to this background and the bubble equations, or integrability conditions, localize the D0 branes at the equator of the $S^2$.  The back-reacted D0 brane gas is, of course, an extremely singular family of solutions, however, it is then argued that the Myers effect \cite{Myers:1999ps} causes  polarization of the D0 branes into D2 branes wrapping the $S^2$. If this configuration is lifted to $M$-theory then the D0-charge becomes momentum and the underlying configuration can be mapped onto the MSW string \cite{Maldacena:1997de} and can thus carry the entropy of a BPS black hole. 

The Myers effect can only produce dielectric D2 branes and so it is something of a `stretch' to get to non-trivial wrapped branes on the $S^2$.  There are also  issues of tadpoles and supersymmetry breaking that we will discuss in more detail below.

Another approach that points to importance of branes wrapping cycles of microstate geometries comes from quiver quantum mechanics \cite{Denef:2000nb,Denef:2002ru,Bates:2003vx,Denef:2007vg}.  The underlying supergravity background is once again a scaling geometry constructed from fluxed D6 branes.  In terms of the quantum mechanics on the brane, it is shown in \cite{Denef:2007vg,deBoer:2008zn,Bena:2012hf,Lee:2012sc,Manschot:2012rx} that the exponentially growing number of states of a black hole can arise only on the Higgs branch of the quiver. The states on the Higgs branch are described by open strings stretched between the D6 branes, and in $M$-theory this corresponds to M2 branes wrapping the non-trivial cycles of the corresponding scaling geometry.

Much more recent work \cite{Martinec:2015pfa} by Martinec and Niehoff provides a new framework that gives  new insights into deconstruction and the ground-state structure  of quiver quantum  mechanics.   One of the key observations in  \cite{Martinec:2015pfa} is that if branes wrap cycles in scaling microstate geometries, then they become very light, or give rise to either massless particles or tensionless strings or branes in the extreme scaling limit.  It is also suggested that such  ``W-branes'' could condense and give rise to new phases with an exponentially growing number of BPS states in those new phases.  In this way one can imagine the  W-brane  degrees of freedom giving rise to the Higgs branch of quiver quantum mechanics.  

The apparent problem with this idea is that, naively, there appear to be  rather few ways that branes can wrap cycles in a microstate geometry.  However, as in the deconstruction story, Martinec and Niehoff point out that within the compactified directions, the wrapped branes behave as point particles in the magnetic fluxes that thread the compact directions and so there is a vast number of distinct W-branes coming from the degeneracy of lowest Landau level. They then argue that the W-brane degeneracy is given by counting walks on the quiver and show how this reproduces the counting formulae derived in \cite{Denef:2007vg,Bena:2012hf,Lee:2012sc,Manschot:2012rx} and hence the exponentially growing number of states.

This is an extremely appealing picture in that it not only meshes well with the ideas and results coming from deconstruction and from quiver quantum mechanics but it also provides a semi-classical description of solitonic states, within a microstate geometry,  whose massless limit may well account for the black-hole entropy.  The picture is also beautifully reminiscent of how the $E_8 \times E_8$ W-bosons emerge from the geometry of $K3$ in heterotic-type II duality \cite{Hull:1994ys}.  The W-brane picture exemplifies how microstate geometries can provide a background to describe intrinsically stringy states that are part of the microstate structure of black holes.

Whatever the perspective, there are  compelling reasons to study branes wrapping cycles in microstate geometries and, in particular, M2 branes wrapping $2$-cycles in the original five-dimensional microstate geometries  \cite{Bena:2005va,Berglund:2005vb,Bena:2007kg} in M-theory.   

Up until recently,  W-branes have been treated as probes in the background of microstate geometries.  However, if there are sufficiently many  wrapped branes then there will be a significant back-reaction and this will require treatment within supergravity.   It is therefore possible that supergravity may, in fact, see some large-scale, coherent aspects of W-branes. Perhaps the simplest approach to investigating such wrapped branes in supergravity is to start with the $T^6$ compactification of $M$-theory and consider an M2 brane wrapping an $S^2$ in a five-dimensional microstate geometry.  Such a brane is point-like in the $T^6$ but one can simplify the problem by smearing over the entire torus and reducing the problem to five-dimensional supergravity.  The smearing also avoids the problem of how to handle the electric field lines on the compactification manifold since it forces all the electric field lines into the space-time.  In  five dimensions, the four-form field strength sourced by such a wrapped brane  is dual to a scalar field and the relevant five-dimensional field theory is $\Neql2$ supergravity coupled to both vector multiplets and hypermultiplets.  Thus far,  the study of microstate geometries from the five-dimensional perspective has largely focussed on $\Neql2$ supergravity coupled only to vector multiplets; the addition of hypermultiplets add a whole new level of complexity but is required if one is to study wrapped brane states in five dimensions.  This has therefore been the starting point for studying the supergravity back-reaction of wrapped branes 
  \cite{Levi:2009az,Raeymaekers:2014bqa,Raeymaekers:2015sba}. 

There is a potential  problem with smearing the brane source over the compactification manifold:  such a source has spatial  co-dimension equal to $2$ and so, in flat space, would involve logarithmic Green functions. In asymptotically-flat backgrounds this will generically lead to singularities at infinity.   One can avoid this issue by taking the background to be $AdS_3 \times S^2$. Indeed, if one takes a bubbled microstate geometry and removes the constants in the harmonic functions that lead to  asymptotically flatness at infinity, one typically gets a solution that is asymptotically $AdS \times S$.  This means that $AdS_3 \times S^2$ is  an excellent local model of a single bubble within a scaling microstate geometry.  If the total W-brane charge is non-zero then restoring the asymptotically flat regions will still lead to  problems.  In an asymptotically flat solution one can interpret the configuration in terms of branes, specifically in the IIA formulation, the wrapped M2 W-branes become fundamental strings ending on the D6 that is wrapped on a $T^6$.  This leaves an uncanceled tadpole on the compact D6 world-volume.  This issue is presumably related to divergences at infinity in the supergravity solution. The easiest way to handle these problems in flat space is to consider a solution with multiple bubbles and try to wrap branes in such a manner that there is no net W-brane charge, leading to a dipolar charge distribution whose fields fall off faster at infinity in supergravity and have no tadpoles on other wrapped branes.  We will see how this is completely incompatible with supersymmetry.

Wrapped M2 branes in $AdS_3 \times S^2$ were studied in great detail in   \cite{Levi:2009az,Raeymaekers:2014bqa,Raeymaekers:2015sba}, where some very interesting new  families of BPS solutions were found.  These new families were shown to preserve $4$ supersymmetries but it was not clear how those supersymmetries would be modified in an asymptotically flat background and how the supersymmetries might depend upon the orientation of one bubble relative to another.  It is the purpose of this paper to resolve precisely these issues.  

First and foremost, if a solution is asymptotic to $AdS \times S$  or, equivalently, the dual field theory is superconformal, then the supersymmetries come in two classes:  Poincar\'e and superconformal.   Breaking the conformal invariance can only preserve the  Poincar\'e supersymmetries.  For a black hole in flat space, the superconformal symmetry present in the near-horizon limit is broken by the flat asymptotics.   In the standard lexicon, when we say that a \nBPS{8} black hole has four supersymmetries, this means four {\it Poincar\'e} supersymmetries.  Moreover, any microstate of such a black hole, and thus any corresponding microstate geometry, must also preserve the same four  {\it Poincar\'e} supersymmetries.  To understand whether the solutions of  \cite{Levi:2009az,Raeymaekers:2014bqa,Raeymaekers:2015sba} lie in the ensemble of \nBPS{8} black hole microstates therefore rests on understanding  how many of the four supersymmetries  identified in  \cite{Levi:2009az,Raeymaekers:2014bqa,Raeymaekers:2015sba} are Poincar\'e or superconformal supersymmetries and whether these solutions, in fact, break the Poincar\'e supersymmetries of the \nBPS{8} black hole.  By analyzing the supersymmetries when these solutions are coupled to flat space, we will show that two of the four supersymmetries are broken by that coupling, which means that there are only two Poincar\'e supersymmetries and thus the solutions of \cite{Levi:2009az,Raeymaekers:2014bqa,Raeymaekers:2015sba}  are  \nBPS{16} states.

To understand the supersymmetry in more detail,  one can study the supersymmetry projectors that arise from the corresponding brane configuartions. {\it A priori}  there are two possible ways in which the supersymmetries may work out.   The first, and most obvious, comes from the standard application of brane projectors:  A stack of M2 branes lying in directions $0,1,2$ impose the following constraint on supersymmetries  (see, for example, \cite{Polchinski:1998rr,Stelle:1998xg}): 
\begin{equation}
\Gamma^{012}  \, \epsilon ~=~ \epsilon\,. \label{naiveproj}
\end{equation}
and thus, typically, cut the amount of supersymmetry in half compared to the amount  of supersymmetry without the M2 branes.  In particular, applying one such projector to a standard \nBPS{8} microstate geometry would result in a \nBPS{16} background. If there were multiple wrapped branes on cycles with different orientations in the four-dimensional spatial base, one would have to impose at least one other projector of the form $\Gamma^{0xy}$, and since this does not commute with (\ref{naiveproj}), all supersymmetry would be broken.

However, the projector  (\ref{naiveproj}) is based  upon rather simple brane configurations and is possible that it is modified in a suitably complicated background.  For example, the dipolar distributions of M5 charge that underlie black rings and microstate geometries neither modifies nor places further conditions on the supersymmetries apart from the projections required by the original electric M2-brane charges.  This works through a remarkable conspiracy between the  M5-charge density and the angular-momentum density so that their combined supersymmetry projections reduce to those of the underlying M2-brane charges \cite{Bena:2004de,Bena:2011uw}.  Thus microstate geometries exhibit the kind of geometric transitions that allow densities of new brane charges in precisely such a way that the original supersymmetries remain unbroken.  However, such remarkable conspiracies can usually be detected by brane probes. Brane wrapping of black-hole geometries was also analyzed in detail in \cite{Das:2005za}, where it was shown, using brane probes, that branes that wrap black holes in asymptotically flat geometries generically break all the supersymmetry.  
  
Here we will show that wrapping  M2 branes on a space-time $2$-cycle does indeed reduce the supersymmetry in exactly the manner that the naive brane-projector analysis suggests.  In particular, we will show that two of the four supersymmetries found in \cite{Raeymaekers:2015sba} are artefacts of the superconformal symmetry and will be lost as soon as the configuration is embedded in an asymptotically-flat space-time.  Indeed,  for \nBPS{8} microstate geometries that are asymptotic to flat space necessarily impose the supersymmetry projection condition:
\begin{equation}
\Gamma^{1234}  \, \epsilon ~=~ \epsilon\,, \label{SDproj}
\end{equation}
where $1,2,3,4$ are the non-compact spatial directions (see, for example \cite{Bena:2006kb}) and the $\Gamma$'s are eleven-dimensional gamma-matrices.  This condition arises from the fact that the configuration must carry three M2-brane charges.  Put differently, to represent a microstate of a black hole, the microstate geometry must have the same supersymmetry of the black hole.  For five-dimensional black holes, where the supersymmetries  are symplectic Majorana, the projection condition  (\ref{SDproj}) may be written:
\begin{equation}
\gamma^{0}  \, \epsilon^i ~=~ i {{\sigma_3}^i}_j   \, \epsilon^j  \,, \label{BHproj}
\end{equation}
where $\gamma^{0}$ is a five-dimensional gamma-matrix.
Using the standard identity for the product of gamma matrices, this may be re-written as (\ref{SDproj}).  We show precisely how imposing this projection condition on the supersymmetries of  \cite{Raeymaekers:2015sba}, cuts their number in half, which means that, once embedded in flat space, these solutions are actually \nBPS{16} microstates.  Going further, we relate the computations in \cite{Raeymaekers:2015sba} to their flat-space analogs, and show how the projection conditions imposed in \cite{Raeymaekers:2015sba} are precisely of the form (\ref{naiveproj}).

Thus we conclude that for a microstate geometry in flat space (where conformal invariance is broken), wrapping branes around a single cycle will reduce the usual \nBPS8, asymptotically flat, microstate geometries, with four (Poincar\'e) supersymmetries, to \nBPS{16} microstate geometries, with just two (Poincar\'e) supersymmetries. Furthermore, the supersymmetry projection that plays an integral role in \cite{Raeymaekers:2015sba} actually depends upon the orientation of the wrapped cycle and if cycles have different orientations then the corresponding supersymmmetry projectors will be incompatible.  This means that wrapping more than one cycle with generic orientations will, in fact, break all the supersymmetries.  We will also show that there is no way to solve the tadpole problem without breaking all the supersymmetries and so we recover the result expected from brane-probe analysis  \cite{Das:2005za}.

In Section 2 give some of the relevant details of $\Neql 2$ supergravity in five dimensions coupled to vector multiplets and hypermultiplets.  In Section 3 we set the hypermultiplets to zero and  review (very briefly)  the standard formulation of bubbled microstate geometries using a Gibbons-Hawking (GH) base for the spatial sections of the manifold and then discuss how $AdS_3 \times S^2$ emerges as a local model of an isolated bubble.  We then discuss the eight supersymmetries of $AdS_3 \times S^2$  in terms of the global metric and in the Bergman form and show how these supersymmetries are reduced to four if the superconformal symmetry is broken by adding more bubbles or simply making the single-bubbled solution asymptotically flat. In Section 4 we restore the hypermultiplets and consider the solutions of  \cite{Raeymaekers:2015sba}.  We discuss the structure of the supersymmetry and how it is further reduced  by the presence of wrapped branes and we translate this back into the description of bubbled geometries using GH  base geometries.  This enables us to show how the supersymmetry will generically be completely broken by wrapping branes on multiple cycles. We argue that the only supersymmetric microstate geometries with wrapped branes are \nBPS{16} and these involve branes wrapped in the same orientation around co-linear cycles.  Moreover, we argue that any solution with  branes wrapped in the space-time and with no net charge for these branes necessarily breaks all the supersymmetry.   Finally, in Section 5 we discuss the meaning of our result for the study of W-branes and black-hole microstates. In particular, we argue that while the wrapped-brane solutions are not, in themselves,  \nBPS8 microstates  of black holes or black rings, W-branes do provide a way to access and might even enable us to  count the BPS microstate structure of \nBPS8 black holes and black rings  in deep scaling microstate geometries.

\section{The Lagrangian and BPS equations}

We work within five-dimensional, $\Neql2$ supergravity coupled to both vector and hypermultiplets.  The bosonic action may be taken to be:
\begin{equation}
\begin{aligned}
  S ~=~  \int\!\sqrt{-g}\,d^5x \Big( R  ~-~ & Q_{IJ} \,\partial_\mu X^I  \partial^\mu X^J~-~ \coeff{1}{2} h_{uv} \, \cD_\mu q^u \cD^\mu q^v ~-~\coeff{1}{2} Q_{IJ} F_{\mu \nu}^I   F^{J \mu \nu}\\ 
  & ~-~\coeff {1}{24} C_{IJK} F^I_{ \mu \nu} F^J_{\rho\sigma} A^K_{\lambda} \bar\epsilon^{\mu\nu\rho\sigma\lambda}\Big) \,.
\end{aligned}
  \label{5daction}
\end{equation}
Our goal is to write the action in a manner that is a simple extension of $\Neql2$ supergravity coupled to  vector multiplets that is typically used in the discussion of microstate geometries.  Our space-time metric is ``mostly plus''  and we will only have two vector multiplets and hence three vector fields.  Thus $I, J =1,2,3$, and we normalize the $A^I$ so that $C_{123} =1$.  The  scalars satisfy the constraint $X^1 X^2 X^3 =1$ and metric for the kinetic terms is: 
\begin{equation}
 Q_{IJ} ~=~    \frac{1}{2} \,{\rm diag}\,\big((X^1)^{-2} , (X^2)^{-2},(X^3)^{-2} \big) \,.
\label{scalarkinterm}
\end{equation}
As usual, it is convenient to introduce three scalar fields, $Z^I$, and take
\begin{equation}
Z  ~\equiv~  (Z_1Z_2 Z_3)^{1/3} \,, \qquad X^J   ~\equiv~ \frac{Z}{Z_J} \,, \qquad X_J   ~\equiv~ \frac{1}{3}\, \frac{Z_J}{Z}     \,.
\label{XZrelns}
\end{equation}
The scalars, $q^u$, are, of course, those of the hypermultiplets. 

One can easily relate our conventions most simply to those of \cite{Bergshoeff:2004kh}.  Define 
\begin{equation}
\hat A^I  ~\equiv~ -\sqrt{3}\,  A^I  \,, \quad \hat C_{IJK}   ~\equiv~\coeff{1}{6} \, C_{IJK}  \,, \quad h^I ~\equiv~ X^I   \,, \quad h_I ~\equiv~ X_I   \,, \quad a_{IJ}  ~\equiv~ \coeff{2}{3} \,  Q_{IJ} \,.
\label{hatAC}
\end{equation}
then the hatted quantities are those of   \cite{Bergshoeff:2004kh}  and we have set $\kappa = \frac{1}{\sqrt{2}}$.  The conventions of \cite{Bellorin:2006yr} are very similar, except they use a ``mostly minus'' metric and thus one must send $g_{\mu \nu} \to - g_{\mu \nu}$ and modify gamma matrices appropriately.

The BPS equations come from setting all the supersymmetry variations of the fermions to zero:
\begin{eqnarray}
 \nabla_\mu \epsilon^i ~+~ \coeff{i}{8}\, X_I  \, {F^I\,}^{\nu \rho} \, \big( \gamma_{\mu \nu \rho} - 4 \, g_{\mu \nu} \gamma_\rho \big) \, \epsilon^i ~-~ \partial_\mu  q^v  \, {\omega_v}^{ij}\epsilon_j &~=~ 0 \,,  \label{deltapsi} \\ 
\Big[ i \, \gamma^\mu \partial_\mu X^I ~+~ \coeff{1}{2} \, \big( \delta^I_J ~-~ X^I X_J\big)\,{F^I}^{\rho \sigma}   \gamma_{\rho \sigma} \Big] \, \epsilon_i &~=~ 0\,,   \label{deltalambda} \\ 
 i \, \gamma^\mu \,(\cD_\mu q^v) \, f_v^{jA} \, \epsilon_j &~=~ 0\,.   \label{deltazeta}
\end{eqnarray}
The symplectic indices are raised and lowered using 
\begin{equation}
 v^i ~=~ \epsilon_{ij} \, v_j  \,, \qquad   v_i ~=~  v^j \, \epsilon_{ji}    \,,
\label{raiselower}
\end{equation}
and our gamma matrices satisfy 
\begin{equation}
\begin{aligned}
\big \{\, \gamma^a \,,\gamma^b \, \big\} & ~=~2\,\eta^{ab} \,, \qquad \gamma^{abcde} ~=~ i \epsilon^{abcde}\,, \qquad \epsilon^{01234} ~\equiv~ +1 \,,\\
(\gamma^0)^\dagger  & = - \gamma^0 \,,\qquad (\gamma^A)^\dagger =  (\gamma^A)^\dagger \,,  \quad A= 1,2,3,4 \,.
\end{aligned}
\label{gammaconv}
\end{equation}
%

\section{The standard bubbled geometries}

\subsection{Bubbled geometries on a Gibbons-Hawking base}

We first set all the hypermultiplet scalars to zero and recall the story for bubbled geometries on a GH base manifold, $\cB$. 
The metric takes the form:
\begin{equation}
ds_5^2 ~\equiv~ - Z^{-2} \big(dt+ \mu (d\psi+A) + \omega\big)^2 ~+~
Z\, \big(   V^{-1} (d\psi+A)^2 ~+~ V(d \vec y \cdot d \vec y)  \big)\,,
\label{fivemetric}
\end{equation}
where the spatial sections on $\cB$ are the usual, possibly ambi-polar, GH metric with 
\begin{equation}
\vec \nabla \times \vec A ~=~ \vec \nabla V\,.
\label{AVreln}
\end{equation}
Later it will be useful  to adopt axial polars on the $\IR^3$ sections of $\cB$, in which one has:
\begin{equation}
d \vec y \cdot d \vec y~=~ d\rho^2 ~+~  \rho^2 d\phi^2 ~+~  dz^2 \,.
\end{equation}
As usual we take:
\begin{equation}
V ~=~ \varepsilon_0 ~+~ \sum_{i =1}^N \, \frac{q_i}{r_i} \,,
\label{Vform1}
\end{equation}
where $r_i \equiv |\vec y - \vec y_i|$.

The BPS Ansatz for the Maxwell fields may be decomposed into 
electric and magnetic components:
\begin{equation}
A^I   ~=~  -  Z_I^{-1}\, (dt +\mu (d\psi+A) + \omega) ~+~ B^{(I)}  \,,
\label{Aform}
\end{equation}
where $B^{(I)}$ is a one-form on $\cB$.   The magnetic parts of the field strengths are defined by:
\begin{equation}
\Theta^{(I)}    ~\equiv~  d B^{(I)}    \,.
\label{Thetadefn}
\end{equation}
The magnetic vector potentials are given by:  
\begin{equation}
B^{(I)}  ~=~  V^{-1}\,   K^I  \, (d\psi ~+~ A) ~+~  \vec{\xi}^{(I)}   \cdot d \vec y \,,  \qquad \vec  \nabla \times \vec \xi^{(I)}    ~=~ - \vec \nabla K^I  \,,
\label{Bpot}
\end{equation}
while the electrostatic potentials are
\begin{equation}
Z_I  ~=~  \coeff{1}{2}\,  C_{IJK}  \, \frac{ K^J \, K^K}{V}  ~+~  L_I\,,
\label{ZIs}
\end{equation}
where
\begin{equation}
K^I  ~=~  \sum_{i =1}^N \, \frac{k^I_i}{r_i} \,, \qquad L_I  ~=~  \ell_0 ~+~ \sum_{i =1}^N \, \frac{\ell_{I\, i}}{r_i}  \,.
\label{KLfns}
\end{equation}
The remaining parts of the metric are given by:
\begin{align}
\mu  ~=~& \frac{K^1 K^2 K^3}{V^2} ~+~ \frac{1}{2}\,    \frac{K^I \, L_I}{V} ~+~ M \,,  \qquad  M  ~=~  m_0 ~+~ \sum_{i =1}^N \, \frac{m_i}{r_i} \,,\label{mueqn}\\
\vec \nabla \times \vec \omega ~=~&  V \vec \nabla M ~-~ M \vec \nabla V ~+~   \coeff{1}{2}\, (K^I  \vec\nabla L_I - L_I \vec \nabla K^I )\,.
\label{omegeqn}
\end{align}
Regularity at each GH then requires that:
\begin{equation}
\ell_j^I  ~=~ -  \coeff{1}{2}\,  C_{IJK} \,  { k_j^J \, k_j^K  \over q_j} \,,  \qquad m_j ~=~  \coeff{1}{2}\,  {k_j^1 \, k_j^2 \, k_j^3 \over q_j^2}  \,, \qquad j=1,\dots, N \,.
\label{lmchoice}
\end{equation}
The parameters $\varepsilon_0$, $\ell_0$ and $m_0$ are determined by the asymptotics at infinity.

Finally, there are the bubble equations that must be satisfied so as to avoid closed timelike curves at each of the GH points.  For more details, see \cite{Bena:2007kg}.

\subsection{Two centers and $AdS_3 \times S^2$}

If one can separate two of the GH centers from the rest and if they are close enough together so that one can ignore the constants, $\varepsilon_0$ and $\ell_0$,  then the resulting space-time may be reduced to $AdS_3 \times S^2$  \cite{Denef:2007yt,deBoer:2008fk,Bena:2010gg}.  The GH potential is simply: 
\begin{equation}
V ~=~ { q_+ \over r_+} ~-~ {q_- \over r_-} \,, 
\end{equation}
where $q_\pm \ge 0$ and
\begin{equation}
r_\pm ~\equiv~   \sqrt{\rho^2 ~+~ (z\mp a)^2 } \,.
\end{equation}
Gauge transformations allow us to shift $K^I  \to K^I + c^I V$, which means we can shift the poles in  the $K^I$ and assume, without loss of generality, that 
\begin{equation}
K^I  ~=~  k^I\, \Big({ 1 \over r_+} ~+~ {1 \over r_-} \Big)   \,.
\end{equation}
By uplifting to six dimensions one can shift $V$ by one of the $K^I$'s and such a spectral flow can be used to set 
\begin{equation}
V~=~ q\, \Big({ 1 \over r_+} ~-~ {1 \over r_-} \Big)  \,, 
\label{Vform2}
\end{equation}

For simplicity, we will take:
\begin{eqnarray}
\label{VK} 
V &=&  q\, \Big({ 1 \over r_+} ~-~ {1 \over r_-} \Big)\,, \qquad   K^I ~=~ K ~=~
k\, \Big({ 1 \over r_+} ~+~ {1 \over r_-} \Big)  \,, \\
\qquad  L_I&=& L ~=~ - {k^2 \over q} \Big({ 1 \over r_+} ~-~ {1 \over r_-} \Big) \,, \qquad  M ~=~
- {2\, k^3 \over a\, q^2}~+~ \frac{1}{2} \,{k^3 \over q^2} \Big({ 1 \over r_+} ~+~
{1 \over r_-} \Big)\,,
\label{LM}
\end{eqnarray}
where the forms of the $L_I$ and $M$ are determined by regularity. One then finds
\begin{eqnarray}
\label{Zmuform}
Z_I  &=& Z  ~=~  V^{-1} K^2 + L ~=~ - {4\, k^2 \over q}  \, { 1 \over (r_+ - r_- )}\,,  \\
\mu &=&  V^{-2} K^3 +  \coeff{3}{2}\, V^{-1} K \, L + M ~=~ {4\, k^3 \over q^2}  \,
{  (r_+ + r_- ) \over (r_+ - r_- )^2} ~-~  {2\, k^3 \over a\, q^2} \,. 
\label{Mu}
\end{eqnarray}
The one forms are given by: 
\begin{equation}
\begin{aligned}
A ~=~ &  ~q\, \Big({(z -a) \over r_+} - {(z +a) \over r_-} \Big) \, d\phi  \,, \qquad  \vec{\xi} \cdot d \vec{y} ~=~  -k\, \Big({(z -a) \over r_+} + {(z +a) \over r_-} \Big) \, d\phi \,, \\
 \omega ~=~ & -{2\, k^3 \over a\, q} \, {\rho^2 + (z-a +r_+)(z+a - r_-)  \over r_+ \, r_-}  \, d\phi  \,.
\end{aligned}
\label{Axiomg}
\end{equation}
This metric (\ref{fivemetric}) is equivalent to the $AdS_3 \times S^2$ space-time and can be written in the global form by performing the following coordinate transformation:   
\begin{equation}
 z =  a\, \cosh 2\xi \,\cos \theta \,, \qquad  \rho =  a\, \sinh 2\xi \, \sin \theta \,, \qquad
 \xi \ge 0\,, \ \ 0 \le \theta \le \pi \,,
  \label{coordsa}
 \end{equation}
and shifting and rescaling variables:
\begin{equation}
 \tau ~\equiv~   \coeff{a\, q}{8\, k^3}\,  t \,, \qquad \varphi_1 ~\equiv~   \coeff{1}{2\,q} \, \psi -
 \coeff{a\, q}{8\, k^3}\,  t   \,, \qquad \varphi_2 ~\equiv~ \phi -   \coeff{1}{2\,q} \, \psi +
 \coeff{a\, q}{4\, k^3}\,  t   \,.
 \label{coordsb}
 \end{equation}
The metric (\ref{fivemetric}) then takes the simple form: 
\begin{equation}
ds_5^2 ~\equiv~ R_1^2 \big[ - \cosh^2\xi \,  d\tau^2 + d\xi^2 +  \sinh^2 \xi \, d\varphi_1^2 \big] ~+~  R_2^2 \big[   d \theta ^2 + \sin^2\theta  \, d\varphi_2^2 \big]  \,,
 \label{AdS3S2}
 \end{equation}
where
\begin{equation}
  R_1~=~  2 R_2 ~=~ 4 k \,.
 \label{Radii}
 \end{equation}

The Maxwell field also dramatically simplifies and (\ref{Aform}) reduces to: 
\begin{equation}
A ~=~  -2k\, \cos \theta \, d\varphi_2    \,, \qquad F~=~ dA ~=~  2k\, \sin \theta \, d\theta \wedge d\varphi_2   \,,
 \label{Asimp1}
 \end{equation}
and so, as one would expect, $F$ is proportional to the volume form on the $S^2$.

\subsection{Other frames and coordinates}

There is another standard form of the metric on $AdS_3$ that will prove useful:  The Bergman form, which describes the metric as a non-trivial time fibration over a non-compact K\"ahler base.  The  $AdS_3$ factor of  (\ref{AdS3S2}) can be written as 
\begin{equation}
ds_3^2 ~\equiv~\frac{ R_1^2}{4}\, \big[ \,- ( d \hat t + 2 \sinh^2 \coeff{1}{2} \zeta\, d\hat \psi )^2 ~+~  d\zeta^2 ~+~   \sinh^2 \zeta \, d\hat \psi^2  \, \big]  \,,
 \label{AdS3Berg}
 \end{equation}
This comes from a very simple change of variable in  (\ref{AdS3S2}):
\begin{equation}
\xi ~=~ \coeff{1}{2}\, \zeta    \,, \qquad \tau ~=~ \coeff{1}{2}\, \hat t    \,, \qquad \varphi_1  ~=~  (\hat \psi  - \coeff{1}{2}\, \hat t ) \,.
 \label{coordchg1}
 \end{equation}
In particular, this and  (\ref{coordsb}) implies
\begin{equation}
\hat \psi  ~=~ \coeff{1}{2\,q}\, \psi  \,.
 \label{coordchg2}
 \end{equation}

It will be convenient to introduce three sets of frames for each of the three forms of the metric:   
\begin{align}
e^0 &~=~ Z^{-1} \,  (dt+\mu(d\psi+A)+\omega) \,,  \qquad e^1 ~=~ Z^{1/2}V^{-1/2} \,  (d\psi+A) \,,  \nonumber \\
e^2 &~=~ Z^{1/2}V^{1/2} \,  d\rho \,,  \qquad  e^3  ~=~ Z^{1/2}V^{1/2}\, \rho \,  d\phi \,, \qquad   e^4  ~=~ Z^{1/2}V^{1/2} \,  dz \,; 
\label{GHframes}\\[8pt]
\tilde e^0 &~=~ 4k\,   \cosh \xi \, d\tau  \,,  \qquad \tilde e^1 ~=~ 4k \,  d \xi  \,, \qquad \tilde e^2 ~=~ 4k\,    \sinh \xi \,d\varphi_1  \,,  \nonumber \\
  \tilde e^3  &~=~2k\, d\theta  \,, \qquad   \tilde e^4  ~=~ -2k\,  \sin\theta\, d\varphi_2 \,;  \label{Globalframes}\\[8pt]
\hat e^0 &~=~ 2 k\,  ( d \hat t + 2 \sinh^2 \coeff{1}{2} \zeta\, d \hat \psi ) \,,  \qquad \hat e^1 ~=~ 2k \,  d \zeta  \,, \qquad  \hat e^2 ~=~ 2k\,    \sinh \zeta\,  d\hat \psi  \,,  \nonumber \\
\hat e^3  &~=~2k\, d\theta  \,, \qquad     \hat e^4  ~=~ -2k\,  \sin\theta\, d\varphi_2 \,.  \label{Bergframes}
\end{align}
The negative sign in $\tilde e^4$ and $\hat e^4$ might seem unusual but it is there to ensure that the Lorentz transformation from the frames (\ref{GHframes}) to (\ref{Globalframes}) or (\ref{Bergframes}) has determinant equal to $+1$.

It is a trivial exercise to verify
\begin{equation}
\hat e^0  ~=~ \cosh \xi \,  \tilde e^0 ~+~  \sinh \xi \, \tilde e^2    \,, \qquad \hat e^2~=~ \sinh \xi\, \tilde e^0 ~+~  \cosh \xi \, \tilde e^2   \,.
 \label{Ltrf1}
 \end{equation}
The Lorentz transform between the $e^a$ and the $\tilde e^a$ is given by: 
\begin{equation}
\begin{aligned}
 e^0 & ~=~\frac{1}{\cos \theta}\, \big[  \cosh \xi   \, \tilde e^0 ~+~ \sinh \xi   \, \tilde e^2  ~+~  \sin \theta   \, \tilde e^4  \, \big]   \,, \\
  e^1 & ~=~  (\cosh \xi \, \sin \eta \, \tan \theta+   \sinh \xi \, \cos \eta ) \, \tilde e^0 ~+~(\sinh \xi \, \sin \eta \, \tan \theta+   \cosh \xi \, \cos \eta )  \, \tilde e^2  ~+~  \frac{\sin \eta}{\cos \theta}   \, \tilde e^4    \,, \\
       e^3 & ~=~  ( \sinh \xi \, \sin \eta -\cosh \xi \, \cos \eta \, \tan \theta ) \, \tilde e^0 ~+~( \cosh \xi \, \sin \eta -\sinh \xi \, \cos \eta \, \tan \theta  )  \, \tilde e^2  ~-~  \frac{\cos \eta}{\cos \theta}   \, \tilde e^4    \,, \\
 e^2 & ~=~\sin  \eta \,  \tilde e^1 + \cos  \eta \, \tilde e^3    \,, \qquad  e^4~=~\cos  \eta \,  \tilde e^1 -  \sin  \eta \, \tilde e^3   \,.
 \end{aligned}
 \label{Ltrf2}
 \end{equation}
where
\begin{equation}
\cos  \eta  ~\equiv~ \frac{\sinh 2\xi  \, \cos \theta }{\sqrt{\cosh^2 2 \xi - \cos^2 \theta}}   \,, \qquad  \sin  \eta ~\equiv~ \frac{\cosh 2\xi  \, \sin \theta}{\sqrt{\cosh^2 2 \xi - \cos^2 \theta}}     \,.
 \label{etadefn}
 \end{equation}
%

\subsection{Killing spinors}

We continue with all the hypermultiplet scalars set to zero.   Since our background obeys the ``floating brane Ansatz"   \cite{Bena:2009fi} the BPS equation  (\ref{deltalambda})  is trivially satisfied as a result of a cancellation between the connection terms and the Maxwell field strengths.  This leaves the equation
\begin{equation}
 \nabla_\mu \epsilon^i ~+~ \coeff{i}{8}\, X_I  \, {F^I\,}^{\nu \rho} \, \big( \gamma_{\mu \nu \rho} - 4 \, g_{\mu \nu} \gamma_\rho \big) \, \epsilon^i ~=~ 0\,,
 \label{deltapsired}
 \end{equation}
which determines how all the supersymmetries depend upon the coordinates.  Indeed, using the $\tilde e^a$ frames  (\ref{Globalframes}) with (\ref{gammaconv}) to write products of three gamma matrices in terms of products of two gamma matrices,  we find the following differential equations:
\begin{equation}
\begin{aligned}
\partial_\tau \epsilon^j  & ~=~ - \partial_{\varphi_1} \epsilon^j  ~=~  \coeff{1}{2} \, \sinh \xi \, \gamma^{01} \epsilon^j ~-~    \coeff{1}{2} \, \cosh \xi \, \gamma^{12} \epsilon^j  \,,  \qquad  \partial_\xi  \epsilon^j  ~=~  \coeff{1}{2} \,  \gamma^{02} \epsilon^j \,,   \\
 \partial_\theta  \epsilon^j   & ~=~ - \coeff{i}{2} \, \gamma^{4} \epsilon^j \,,      \qquad   \partial_{\varphi_2} \epsilon^j   ~=~ - \coeff{1}{2} \, \cos \theta \, \gamma^{34} \epsilon^j ~-~    \coeff{i}{2} \,  \sin \theta \, \gamma^{3} \epsilon^j    \,.
 \end{aligned}
 \label{AdS-S-BPS}
 \end{equation}
One can trivially solve for the dependence on $\xi$ and $\theta$ and the rest can be solved direct by taking derivatives and commuting gamma matrices through the first part of the solution. We find 
\begin{equation}
 \epsilon^j ~=~ e^{\frac{1}{2} \, \xi \, \gamma^{02} }\, e^{-\frac{i}{2} \, \theta \, \gamma^{4} }\, e^{\frac{1}{2} \, (\varphi_1 - \tau)  \, \gamma^{12} }\,e^{-\frac{1}{2} \, \varphi_2 \, \gamma^{34} }\,\epsilon^j_0 \,.
 \label{AdS-S-KSp1}
 \end{equation}
where $\epsilon^j_0$ is a constant spinor.  Note that there are eight solutions:  four components and two choices for $j$.  These solutions contain both the Poincar\'e and superconformal supersymmetries.

If one uses  the $\hat e^a$ frames  (\ref{Bergframes}) then the local Lorentz rotation  (\ref{Ltrf1}) undoes the $\xi$-dependence and gives 
\begin{equation}
 \epsilon^j ~=~  e^{-\frac{i}{2} \, \theta \, \gamma^{4} }\, e^{\frac{1}{2} \, (\varphi_1 - \tau)  \, \gamma^{12} }\,e^{-\frac{1}{2} \, \varphi_2 \, \gamma^{34} }\,\epsilon^j_0 \,.
 \label{Berg-KSp}
 \end{equation}

Based on (\ref{Ltrf2}), define the ``gamma matrix:''
\begin{equation}
\Gamma^0  ~=~\frac{1}{\cos \theta}\, \big[  \cosh \xi   \,  \gamma^0 ~+~ \sinh \xi   \, \gamma^2  ~+~  \sin \theta   \, \gamma^4  \, \big]    \,.
  \label{Gamma0}
 \end{equation}
Observe that if we take the $\gamma^a$ in this expression to be that gamma matrices in the  $\tilde e^a$ frames in (\ref{Globalframes}), then (\ref{Ltrf2}) implies that $\Gamma^0$ represents the $\gamma^0$ matrix of the GH frames, (\ref{GHframes}).  The natural Poincar\'e projection condition in a generic GH space is given by taking $\Gamma^0  \epsilon^j = \pm i  \epsilon^j$ for one choice of sign.  Acting with $\Gamma^0$ on the spinor in  (\ref{AdS-S-KSp1}) gives:
\begin{equation}
\Gamma^0\, \epsilon^j ~=~  e^{\frac{1}{2} \, \xi \, \gamma^{02} }\, e^{-\frac{i}{2} \, \theta \, \gamma^{4} }\, e^{\frac{1}{2} \, (\varphi_1 - \tau)  \, \gamma^{12} }\,e^{-\frac{1}{2} \, \varphi_2 \, \gamma^{34} }\,\big[ \gamma^0 + \tan \theta \, (i \gamma^0 +\oneone) \,\big]\epsilon^j_0 \,.
 \label{Pproj1}
 \end{equation}
This implies that 
\begin{equation}
\Gamma^0\, \epsilon^j ~=~ i  \epsilon^j  \qquad \Leftrightarrow  \qquad \gamma^0\, \epsilon^j_0  ~=~ i  \epsilon^j_0   \,,
 \label{Pproj2}
 \end{equation}
but that the solution space does not respect the projection with the opposite sign.  The projection condition  (\ref{Pproj2}) therefore identifies the Poincar\'e supersymmetries associated with the general GH space. Note that this is normally recast using (\ref{gammaconv})  so as to emphasize the hyper-K\"ahler property of the base:
\begin{equation}
\Gamma^{1234}\, \epsilon^j ~=~   \epsilon^j  \qquad \Leftrightarrow  \qquad \gamma^{1234}\, \epsilon^j_0  ~=~  \epsilon^j_0   \,.
 \label{Pproj3}
 \end{equation}

Alternatively, based on (\ref{Ltrf1}) and (\ref{Ltrf2}) one can take the $\gamma^a$  to be those of the  Bergman frames, (\ref{Bergframes}), and define the ``gamma matrix:''
\begin{equation}
\widehat  \Gamma^0  ~=~\frac{1}{\cos \theta}\, \big[   \gamma^0   ~+~  \sin \theta   \, \gamma^4  \, \big]  \,.
  \label{Gamma0hat}
 \end{equation}
This is representative of the $\gamma^0$ matrix of the GH frames, (\ref{GHframes}), in the Bergman frames and, acting on  (\ref{Berg-KSp}), it leads to the same result as in (\ref{Pproj2}) and (\ref{Pproj3}).

\section{The supersymmetries with hypermultiplet scalars}

\subsection{The hypermultiplet solutions}

The background considered in \cite{Raeymaekers:2015sba} is the half-hypermultiplet parametrized by a complex scalar, $\tau$.  The simplest way to satisfy the BPS equations for this is to take $\tau = \tau(z)$ to be a holomorphic function of the coordinate, $z  = \tanh \frac{\zeta}{2} \, e^{i\hat  \phi}$
on the Bergman base in (\ref{AdS3Berg}).  Indeed, the simplest non-trivial solution has: 
\begin{equation}
\tau ~=~ - i q_* \, \ln(z) ~+~ i V_{\infty} ~=~  q_* \, \hat \psi ~+~ i (V_{\infty} -  q_* \,\ln ( \tanh \coeff{\zeta}{2})) \,, \label{tausol}
\end{equation}
where $q_*$ and $V_{\infty}$ are constants.    This locates the wrapped M2 branes at $z=0$ with a source proportional to $q_*$.  In \cite{Raeymaekers:2015sba} the solution is written in terms of coordinates $(x,\psi)$ where:
\begin{equation}
\log z ~=~ x + i  \hat \psi    \qquad  \Rightarrow \qquad  \zeta  = - \log\Big(  \tanh\Big(- \frac{x}{2}\Big) \Big)   \,. \label{coords1}
\end{equation}

The new class of solutions obtained in \cite{Raeymaekers:2015sba} have a metric with frames
\begin{equation}
\begin{aligned}
E^{\hat t} &~=~ \frac{l}{2} \,  (d\hat t-(1+\Phi'(x))d\hat \psi   )  \,,  \qquad   E^{\hat x} ~=~  \frac{l}{2}\,  \sqrt{\tau_2} e^{-\Phi(x)} \, dx  \,, \qquad
E^{\hat{\psi}} ~=~ \frac{l}{2} \, \sqrt{\tau_2} e^{-\Phi(x)} \, d\hat \psi    \,, \\
E^{\hat{\theta}} &~=~ \frac{l}{2} \,  d\theta \,, \qquad E^{\hat{\phi}} ~=~ -\frac{l}{2} \, \sin\theta \, (\,d\phi-d\hat \psi + d \hat t \, )~=~ -\frac{l}{2} \, \sin\theta \, d\varphi_2 \,. \\
\end{aligned}
\label{RVframes}
\end{equation}
where $\Phi(x)$ satisfies a non-linear, ordinary differential equation.

Observe that we have made some changes of notation and convention compared to \cite{Raeymaekers:2015sba}.  First, we have relabelled $t$ and $\psi$ in  \cite{Raeymaekers:2015sba} by $\hat t$ and $\hat \psi$.  This makes the notation in (\ref{RVframes}) consistent with our notation here and  avoids the confusion between Im$(\log(z))$ in (\ref{coords1}) and the GH fiber coordinate, $\psi$.  These coordinates are, of course, related by (\ref{coordchg2}).  We have also reversed the sign of $\hat t$ relative to $t$ in  \cite{Raeymaekers:2015sba} and flipped the orientation of the frame $E^{\hat{\phi}}$.  This brings  (\ref{RVframes})  into line with the orientations of (\ref{GHframes})--(\ref{Bergframes}).  It should be remembered that  \cite{Raeymaekers:2015sba} uses conventions that make the two-forms in the BPS equations of bubbled geometries have the opposite dualities to the standard ones on the GH base.  Our modifications restore the canonical forms of these duality conditions.

If one sets $q_* =0$ and thus removes the wrapped M2 branes then one has \cite{Raeymaekers:2015sba}:
\begin{equation}
\Phi(x)=\tilde{\Phi}(x)+\frac12\log V_{\infty} \,, \quad  \tilde{\Phi}(x)=\log(\sinh(-x)) \,, \quad \tau_2 = V_{\infty}
\end{equation}
and one finds that the frames (\ref{RVframes}) become precisely the Bergman frames in  (\ref{Bergframes}) with $l=4k$.

\subsection{The supersymmetries}

 In \cite{Raeymaekers:2015sba} it was shown that non-trivial half-hypermultiplet background imposes one additional projection condition on the supersymmetries if the $AdS_3 \times S^2$ background without the wrapped M2 branes.  This condition is:
\begin{equation}
\big( \oneone \, {\delta^i}_j  ~-~ i \gamma^{\hat{x}\hat{\psi}} \, {{\sigma_3}^i}_j  \big)  \, \epsilon^j   ~=~  0  \,,
\label{proj-hypers}
\end{equation}
where $\sigma_3$ is the usual $2\times 2$ Pauli spin matrix acting on the $\Neql2$ indices of the spinor and $\gamma^{\hat{x}\hat{\psi}}$ refers to the product of gamma matrices in the frames  (\ref{RVframes}).  

We first note that this is a projector in the Bergman basis and so must be applied to the Killing spinor  (\ref{Berg-KSp}).  In particular,  $\gamma^{\hat{x}\hat{\psi}} =  \gamma^{12}$ commutes with all the exponentials in  (\ref{Berg-KSp}) and thus implies: 
\begin{equation}
\big( \oneone \, {\delta^i}_j  ~-~ i \gamma^{\hat{x}\hat{\psi}} \, {{\sigma_3}^i}_j  \big)  \, \epsilon_0^j   ~=~  0  \,,
\label{proj-hypers2}
\end{equation}

Next we observe that (\ref{gammaconv}) implies that 
\begin{equation}
  i \gamma^{\hat{x}\hat{\psi}}    ~=~  - \gamma^{\hat{t}\hat{\theta}\hat{\phi}}     \,,
\label{proj-hypers3}
\end{equation}
and so (\ref{proj-hypers}) is precisely the projector of a brane wrapping the $S^2$. 

Using (\ref{Ltrf1}) and (\ref{Ltrf2}) one can also easily map this projector into standard GH form.  To do this we note that
\begin{equation}
\begin{aligned}
E^{\hat{x}} \wedge E^{\hat{\psi}}   & ~=~  \hat e^1 \wedge  \hat e^2   ~=~  \tilde e^1 \wedge( \sinh \xi\, \tilde e^0+ \cosh \xi \, \tilde e^2)   ~=~  (\sin \eta \, e^2   + \cos \eta \, e^4)  \wedge  (\cos \eta \, e^1   + \sin \eta \, e^3 ) \\ 
&~=~ -\cos^2 \eta \, e^1 \wedge e^4 +  \sin^2 \eta \, e^2 \wedge e^3  -   \sin \eta \,  \cos \eta \, (e^1 \wedge e^2 +e^3 \wedge e^4 )\,,
\end{aligned}
\label{twoformtrf}
\end{equation}
This means that in transforming from the Bergman basis to the GH basis, we have 
\begin{equation}
 \gamma^{\hat{x}\hat{\psi}}    ~\to~   
 -\cos^2 \eta \, \gamma^{14}  +  \sin^2 \eta \, \gamma^{23}    -   \sin \eta \,  \cos \eta \, (\gamma^{12}   + \gamma^{34})\,.
\label{projtrf1}
\end{equation}
However, because of the self-duality of the GH base and the projection condition (\ref{Pproj3}) in the GH frames, we have 
\begin{equation}
 \gamma^{ab} \, \epsilon_{GH}^j   ~=~ - \coeff{1}{2}\, \epsilon^{abcd}\gamma^{cd} \, \epsilon_{GH}^j   \,, 
\label{sdconstr}
\end{equation}
and so (\ref{projtrf1}) becomes
\begin{equation}
i \gamma^{\hat{x}\hat{\psi}}    ~\to~   i  \, \gamma^{23}   ~=~ - \gamma^{014} \,,
\label{projtrf2}
\end{equation}
where we have used (\ref{gammaconv}) in the last identity.  Now recall that the homology cycles in a GH metric are defined by the $\psi$-circle fibered along any curve between poles of $V$.  Moreover, the minimum area cycle involves the shortest such curve.  Thus, in the GH form of the metric with  (\ref{Vform2}), the two cycle is defined by the $\psi$-circle over the interval along the $z$-axis between $-a$ and $a$.  From (\ref{GHframes}), the area form of this cycle is $e^1 \wedge e^4$.  Thus (\ref{projtrf2}) corresponds to the projector for the M2 brane wrapping this cycle.

In a general bubbled solution,  each wrapped M2 brane will give rise to a supersymmetry projector that depends on the orientation of the brane. More  precisely, the supersymmetry projector will depend upon the orientation of the straight line joining the two GH points in the base  $\IR^3$ parametrized by $\vec y$ in (\ref{fivemetric}).   The  $\gamma^{4}$ in (\ref{projtrf2}) will be then replaced by a linear combination of $\gamma^{a}$, $a= 2,3,4$.  Any two such  projectors are compatible (have a common null space) if and only if all the GH points are co-linear and the wrapped branes have the same orientation. Indeed, co-linear wrapped branes with opposite orientations source the Maxwell field with opposite signs and so lead to opposite signs in (\ref{proj-hypers2}). A pair of such opposed projectors manifestly have no common null space.

This has several important consequences for the supersymmetry.  First, {\it all the supersymmetry will be broken if the  branes wrap  cycles that are not co-linear.}  If the wrapped cycles are all co-linear then supersymmetry will still be broken if the branes wrap in different orientations, determined by the relative signs of the Maxwell fields they source.  This means that {\it solutions with wrapped M2 brane but no net wrapped M2-brane charge necessarily break all the supersymmetries.}  Finally, if all the wrapped branes lie on co-linear cycles and have the same orientation then the projectors of these branes are all the same and the combined effect is that they reduce the supersymmetry by another factor of a half. 

As regards the total number of supersymmetries, the $AdS_3 \times S^2$ starts with eight real supersymmetries once the symplectic Majorana condition is imposed on  (\ref{AdS-S-KSp1}) or   (\ref{Berg-KSp}).  If one simply wraps the $S^2$, one preserves the conformal invariance and hence the 
superconformal supersymmetries but one must impose the projector (\ref{proj-hypers2}), and, as shown in \cite{Raeymaekers:2015sba},  this leaves four supersymmetries.    If one  breaks the conformal invariance by either restoring the asymptotically flat region or by adding more bubbles then one must impose another supersymmetry projector,  (\ref{Pproj2}) or (\ref{Pproj3}), which is compatible with the projector (\ref{proj-hypers2}).  This reduces the solution to two supersymmetries, and renders it a \nBPS{16} background.

Thus, if one takes a general \nBPS{8} bubbled geometry and wraps any single bubble with M2 branes, the result is a \nBPS{16} solution. If one wraps more than one bubble then all the supersymmetry will be broken unless all the wrapped bubbles are co-linear and are wrapped in the same orientation and only then  will it  be \nBPS{16}.

\section{Conclusions}
\label{sec:conclusion}

We have shown that wrapped branes do indeed break some, or all, of the supersymmetries in a microstate geometry and that this is governed by the naive supersymmetry projectors associated with the wrapped brane.  Thus branes wrapped on a single cycle, as in    \cite{Levi:2009az,Raeymaekers:2014bqa,Raeymaekers:2015sba}, should really be viewed as a  \nBPS{16} excitations of a microstate geometry because that is the amount of supersymmetry that will survive once superconformal invariance is broken.  This means that such wrapped branes should not be identified with microstates of \nBPS{8} black holes but should be viewed as (partial) supersymmetry-breaking excitations of such geometries.  

Our results also suggest that there might well be interesting \nBPS{16} generalizations of the results in \cite{Raeymaekers:2014bqa,Raeymaekers:2015sba} in which the branes wrapped, with the same orientation, on multiple, co-linear $2$-cycles on a K\"ahler base.  The starting point for such a set of solutions might be to generalize the GH base geometry to the K\"ahler bubbled geometries of LeBrun \cite{LeBrun:1991} and perhaps try to find BPS bubbled solutions in which one adds hypermultiplets to the work of \cite{Bobev:2011kk,Bobev:2012af}.  

On the other hand, one cannot solve the tadpole problem supersymmetrically without decompactifying the compact directions.  To remove the tadpoles, the M2 branes wrapped on space-time cycles must have no net charge.  This can be achieved by wrapping branes around cycles in a closed quiver but this would involve multiple, incompatible supersymmetry projectors.  One might try to use co-linear cycles wrapped in opposite orientations but this would lead to projectors with
\begin{equation}
\Gamma^{012}  \, \epsilon ~=~ \pm \epsilon\,, \label{projpm}
\end{equation}
where the sign depends on orientation, and so there would be no residual supersymmetry.  Therefore, any wrapped brane configuration that solves the tadpole problem without decompactification will necessarily break all the supersymmetries.

The fact that wrapped branes do not  preserve  the supersymmetries of a given microstate geometry means that they should not be viewed as  supersymmetric microstates.  However, this does not mean that they cannot be used to describe the microstate structure.  Indeed, we believe that W-branes can access the \nBPS{8} structure of black holes and that counting W-brane configurations will enable one to enumerate the ground-state degeneracy of \nBPS{8} black holes.

First, Martinec and Niehoff \cite{Martinec:2015pfa} point out that the fact that W-branes are becoming light in the scaling limit means that there will be a new phase of stringy physics emerging in the deep scaling regime of microstate geometries.  They argue that the W-branes will probably form condensates and new operators will develop vevs and define order parameters in that new phase.  Quiver quantum mechanics confirms this picture rather nicely:  when the D6 branes are widely separated, the W-branes are massive Higgs excitations of the system.  When the branes coincide, the Higgs fields become massless and the Higgs branch opens up. The ground-state  degeneracy determined in  \cite{Denef:2007vg,deBoer:2008zn,Bena:2012hf,Lee:2012sc,Manschot:2012rx} amounts to counting all the different vacua on that Higgs branch.   Thus W-branes are one-particle excitations on a new massless branch of physics that is opening up in deep scaling solutions.  To think of W-branes as microstates is to confuse particle excitations with condensates and ground states.   

A simple toy-model is, perhaps, helpful here. Consider the $\cN=2$ Landau-Ginzburg theory in $1+1$ dimensions with superpotential $W = x^{n+2} + a x$, where $x$ is the complex Landau-Ginzburg field and $a$ is a parameter. This model has a residual discrete  $\ZZ_{n+1}$ $\cR$-symmetry.  The F-term constraint shows that there are $n+1$ vacua (Ramond ground states) preserving $\cN=2$ supersymmetry. Between each pair of  vacua, there are minimum energy \nBPS{2} kinks, or solitions carrying a discrete $\cR$-charge.  Individual solitons thus preserve  only $\cN=1$ supersymmetry and multi-soliton states break all the supersymmetry. (See, for example, \cite{Fendley:1990zj}.)  The limit $a\to 0$ corresponds to the $n^{\rm th}$  $\cN=2$ superconformal minimal model \cite{Vafa:1988uu,Martinec:1988zu,Lerche:1989uy}. At the conformal point the Ramond vacuum has an $(n+1)$-fold degeneracy and it is the chiral primary fields that interpolate between these states.   In the limit $a \to 0$, the solitons become massless and are related to combinations of left-moving and right-moving chiral primaries.  For $a \ne 0$, the \nBPS{2} solitons reflect the fundamental degrees of freedom of the massless field theory and yield information about its ground state structure.  In this sense,  W-branes in microstate geometries are, relative to the supersymmetry of the microstate geometry, \nBPS{2} excitations that reflect the new massless degrees of freedom and the ground state structure that will emerge in the deep scaling limit.  Thus, while  $W$-branes are not the \nBPS{8} microstates of the black hole, they do reveal  degrees of freedom that will play an essential role in accessing a large component of the microstate structure.

Another important aspect of the supergravity approach taken in \cite{Levi:2009az,Raeymaekers:2014bqa,Raeymaekers:2015sba} is the smearing of the branes on the  compactification manifold that reduces the problem to five dimensions.  Ignoring the degeneracy of states in the lowest Landau level will, of course, do huge violence to to the state counting and make vast families of distinct W-branes look exactly the same in supergravity.  Indeed, it will collapse the W-brane states to simply the number of ways of wrapping non-trivial cycles in the space-time. In the field theory on the branes this would wash out most of the interesting structure of the new phase that emerges in the deep scaling limit.   It would therefore be extremely interesting to see how one might describe distinct W-branes without smearing in supergravity.  This will almost certainly mean working with higher-dimensional  supergravity theories and finding ways of modeling the distinct Landau orbits or states within the lowest Landau level.    
  
\vspace{5mm}
\noindent
{\bf Acknowledgments}.  
%
We would like to thank Iosif Bena, Emil Martinec, Samir Mathur and David Turton for discussions and we are very grateful to the IPhT, CEA-Saclay for hospitality while this paper was completed.  This work was supported in part by the DOE grant DE-SC0011687.
%

%


\end{document}